\documentstyle[epsfig, twocolumn]{iso98}

\newcommand{\thepapername}{}
\def\etal   {{\it et~al.}}

\begin{document}

\setlength{\parindent}{0pt}
\setlength{\parskip}{ 10pt plus 1pt minus 1pt}
\setlength{\hoffset}{-1.5truecm}
\topmargin -0.3in 
\setlength{\textwidth}{ 17.1truecm }
\setlength{\columnsep}{1truecm }
\setlength{\columnseprule}{0pt}
\setlength{\headheight}{12pt}
\setlength{\headsep}{20pt}
\pagestyle{veniceheadings}

\title{\bf THE FAR-INFRARED CONTINUUM OF QUASARS.}

\author{{\bf B.J.~Wilkes$^1$, E.J.~Hooper$^1$, K.K.~McLeod$^2$,
M.S.~Elvis$^1$, C.D.~Impey$^3$, C.J.~Lonsdale$^4$,} \\
{\bf M.A.~Malkan$^5$,
J.C.~McDowell$^1$} \vspace{2mm} \\
$^1$Smithsonian Astrophysical Observatory, 60 Garden St., Cambridge, MA 02138, USA\\
$^2$Astronomy Department, Wellesley College, Wellesley, MA 02481, USA. \\
$^3$Steward Observatory, University of Arizona, Tucson, AZ 85721, USA \\
$^4$IPAC, Caltech, Pasadena, CA 91125, USA \\
$^5$Department of Astronomy, University of California, Los Angeles, CA
90095, USA}

\maketitle

\begin{abstract}

ISO provides a key new far-infrared window through which to observe
the multi-wavelength
spectral energy distributions (SEDs) of quasars and active galactic
nuclei (AGN). It allows us, for the first time, to observe a substantial
fraction of the quasar population in the far-IR, and to obtain simultaneous,
multi-wavelength observations from 5--200\,$\mu$m. With these data we can
study the behavior of the IR continuum in comparison with expectations
from competing thermal and non-thermal models. A key to determining
which mechanism dominates, is the measurement of the peak wavelength of the
emission and the shape of the far-IR--mm turnover. Turnovers which are
steeper than $\nu^{2.5}$ indicate thermal dust emission in the far-IR.

Preliminary results from our ISO data show broad, fairly smooth, IR
continuum emission with 
far-IR turnovers generally too steep to be
explained by non-thermal synchrotron emission.
Assuming thermal emission throughout leads to a wide inferred temperature range
$\sim50-1000$K.
The hotter material, often called the AGN component,
probably originates in dust close to and heated by the
central source, {\it e.g.} the ubiquitous molecular torus. The cooler
emission is too strong to be due purely to cool, host
galaxy dust, and so indicates either the presence of a starburst in
addition to the AGN or AGN-heated dust covering a wider range of
temperatures than present in the standard, optically thick torus models.
\vspace {5pt} \\  

  Key~words: ISO; infrared astronomy; quasars{}.

\end{abstract}

\section{INTRODUCTION AND ANALYSIS}
The energy output of quasars and AGN is comparable 
thoughout the far-IR to X-ray spectral regions, so studies of their
emission mechanisms require observations throughout this range.
Prior to ISO, the far-IR continuum was observed by IRAS, providing our
first look at quasars in this wavelength range and showing them to be strong
emitters from 12--100\,$\mu$m. The poor spatial resolution and limited
wavelength range of IRAS restricted observations to the IR-brightest subsets
of quasars and AGN and left open many questions as to the range of behavior
present and the emission mechanisms responsible in the various types of
AGN. With ISO we are now able to address some of these limitations by
studying a large number of different kinds of AGN and quasars.

We observed 72 quasars and AGN covering a full range of
redshift and luminosity and many different types of SEDs.
The observations were made with ISOPHOT (\cite{phot96}) and
cover 5--200\,$\mu$m in 8 bands. The status of this
project, including a description of the sample and global comparisons of
our results with those of IRAS, is summarised elsewhere (Hooper \etal ,
this volume). Here we discuss our preliminary results in
the context of the scientific questions we wish to address, with
particular emphasis on the high-redshift objects in the sample. 

The data were processed with PIA (\cite{Gabriel97})\footnote{PIA is a joint
development by the ESA Astrophysics Division and the ISOPHOT Consortium}
through the AAP level, from which custom
scripts (originating with M. Haas \& S. Mueller at MPIA) were used to
determine source fluxes and uncertainties.  Default vignetting values from
PIA were applied to the chopped data, and the flux scale was set by FCS
observations.  Sky noise estimates based on \cite{herb98}
were included in the error budget.
For further details of the data reduction please refer to Hooper \etal\ (this
volume).

\section{SPECTRAL ENERGY DISTRIBUTIONS (SEDs).}
\begin{figure}[h]
    \leavevmode
  \centerline{\epsfig{file=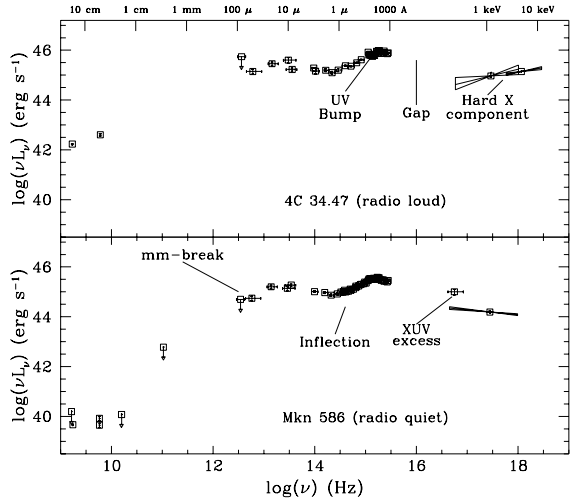,height=3.8truein,angle=-90}}
  \caption{\em Rest-frame radio--X-ray spectral energy distributions (SEDs) for
low-redshift radio-loud (upper, RLQ) and radio-quiet (lower, RQQ)
quasars (Fig.1 from \protect\cite{ewm94}). These are typical
examples of SEDs in the
current literature. The display of $\nu$L$_{\nu}$ vs $\nu$ highlights the
energy output and the structure of the SED. }
  \label{fig:seds}
\end{figure}

Figure~\ref{fig:seds} shows typical examples of the SEDs of a radio-loud
(upper, RLQ) and radio-quiet (lower, RQQ) quasar (\cite{ewm94}).
Throughout the far-IR--ultraviolet (UV) spectral region, the SEDs are
remarkably similar, showing the big blue bump in the optical-UV, the
broad IR bump, and an inflection between the two centered at $\sim 1$\,$\mu$m.
The differences between the two radio classes are evident in the radio
region where, by definition, the RLQs are several orders of
magnitude more luminous, and in the soft X-ray region where the RLQs are
a factor of $\sim 3$ brighter and have a harder spectrum. This latter
difference is generally ascribed to an additional synchrotron
self-Compton component linked to the radio emission (see~\cite{Wilkes99}
for a review of quasar SEDs).

\begin{figure}[h]
    \leavevmode
\centering
\vskip -0.2in
  \centerline{\epsfig{file=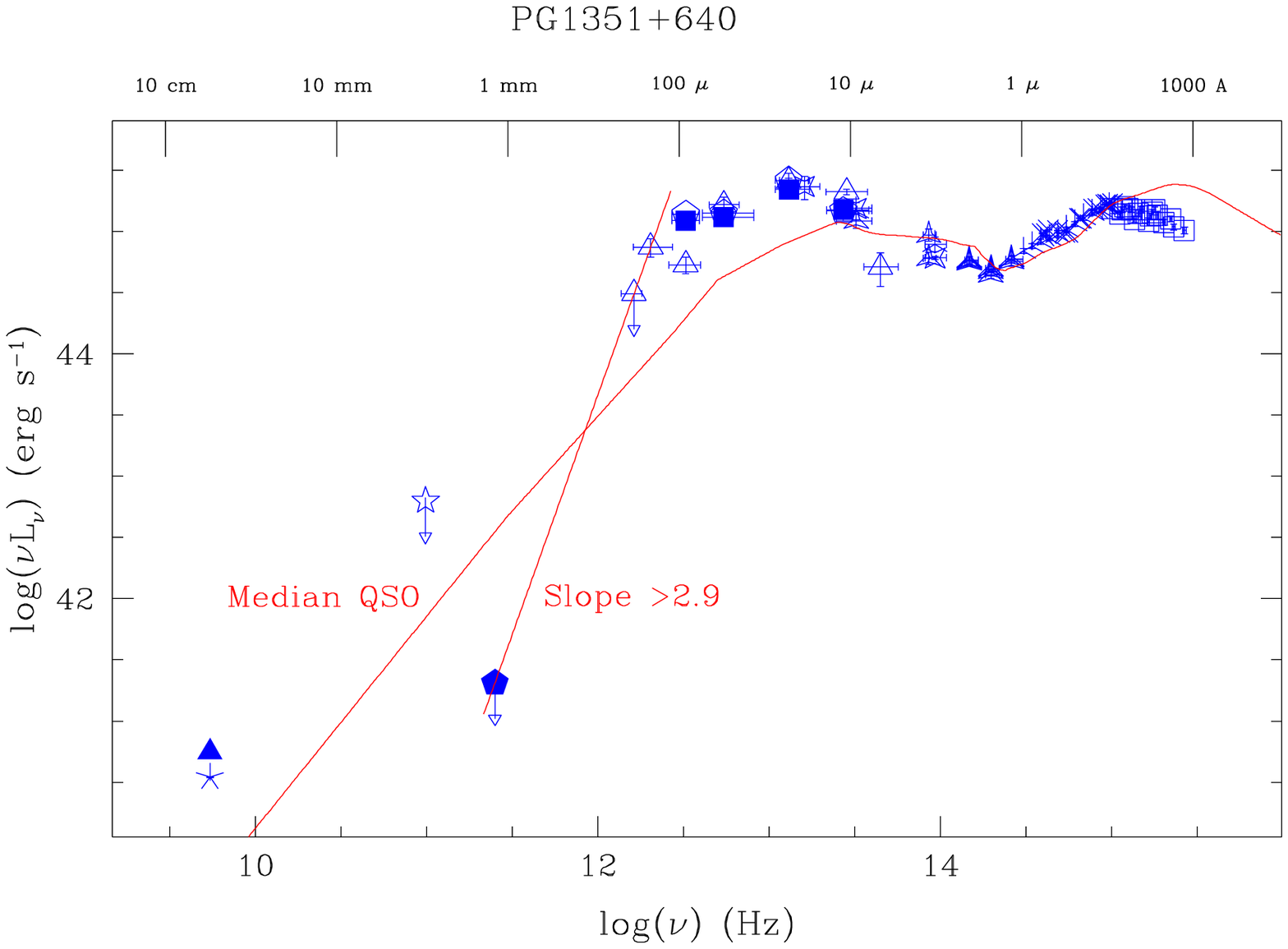,height=3.5truein}}
\vskip -1.2in
  \centerline{\epsfig{file=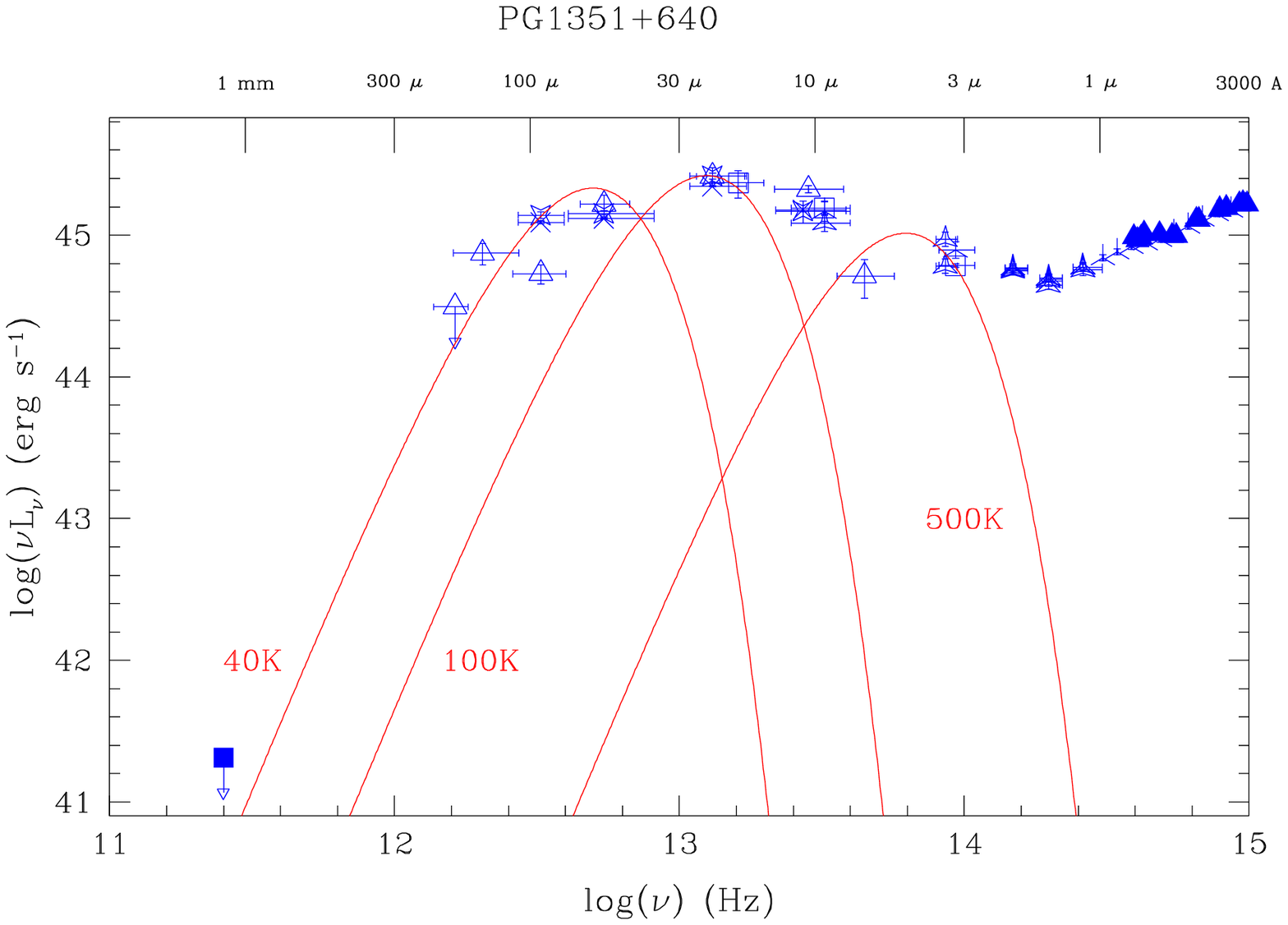,height=3.5truein}}
\vskip -1.0in
  \caption{\em The rest-frame SED of the low-redshift (z=0.088), RQQ
PG1351+640 from \protect\cite{ewm94} with the ISO data points (open triangles)
added. {\bf Upper}: the radio--UV SED showing a 
steep ($\alpha_{fir} > 2.9$) far-IR cut-off;
{\bf Lower}: the IR--optical SED
with representative grey body (assuming $\beta = 2$)
curves superposed.
}
  \label{fig:1351}
\end{figure}

The radio--IR SEDs of core-dominated RLQs are generally smooth
while in both RQQs and lobe-dominated RLQs
the far-IR cutoff is steep and does not extrapolate into the radio.
This difference has resulted in the general belief that the IR emission in
core-dominated RLQs is dominated by beamed, non-thermal and
in RQQs by thermal emission mechanisms (see~\cite{Wilkes99} for a
review). The median RLQ SED in
Figures~\ref{fig:1351},\ref{fig:1202},\ref{fig:1946}
was generated from a combination of lobe- and
core-dominated RLQs. In a unification scenario
(see~\cite{Barthel99} for a review) the beamed IR component
is expected to be weaker in lobe-dominated sources where the beam is in
the plane of the sky. Thus it may not be surprising that the RLQ SED,
formed by combining
SEDs from the two radio subclasses, is similar to that of
the RQQs. A detailed comparison of the SEDs of core- and
lobe-dominated RLQs is required to confirm this scenario. 
Such a transition from non-thermal to thermal emission with decreasing
core-dominance is demonstrated
in the preliminary ISOPHOT results of the European Core Program on Quasars
(\cite{Haas98}; see also this volume). 

The main questions we wish to address with our ISO sample are (i) to better
understand the interplay between
thermal and non-thermal emission in the various types of source (ii)
to investigate the
relative contributions of cool (galactic/starburst) and warm (AGN-related)
dust (iii) to characterize and measure the continua in order to
constrain models (iv) to study the evolution of the IR SEDs.

\subsection{LOW REDSHIFT QUASAR: PG1351+640}
\label{sec:1351}

Figure~\ref{fig:1351} shows the rest-frame SED of a low-redshift (z=0.088),
RQQ PG1351+640. The ISO data points are shown as open triangles and the
remainder of the multi-wavelength data are from~\cite{ewm94}.
This SED is typical of the low-redshift RQQs in our sample and
compares well with the low-redshift median.
The current limit on the
far-IR turnover is $\alpha_{fir} > 2.9$, too steep to be due to
non-thermal synchrotron emission, for which $\alpha_{fir} < 2.5$ is required
and $\alpha < 1$ is more typical (\cite{Gear94}). We thus conclude that
the far-IR emission in this quasars is dominated by thermal emission from dust.
\begin{table*}[t]
\centering
\caption{The Broad-band IR luminosities of the High redshift Quasars.}
\label{tab:lum}
\footnotesize
\begin{tabular}{lllllc}
\hline
Name & Redshift & L(7--25\,$\mu$m)$^1$ & L(2--25\,$\mu$m)$^1$ & L(2--200\,$\mu$m)$^1$ &
\# ISO det'ns \\
\hline
1107+48 & 3.010 & 1.7E+47 & 6.9E+47 & - & 2 \\
BR 1202$-$0727$^2$ & 4.690 & 1.7E+48 & 8.5E+48 & 8.9E+48 & 3 \\
1407+265 & 0.944 & 7.5E+46 & 1.2E+47 & - & 2\\
1413+1143 & 2.551 & 8.8E+47 & 1.6E+48 & 2.0E+48 & 7\\
1718+481 & 1.084 & 2.7E+46 & 8.2E+46 & - & 3\\
HS 1946+7658 & 3.030 & 4.2E+47 & 1.1E+48 & 2.1E+48 & 4 \\
3C422 & 0.942 & 7.6E+46 & 1.4E+47 & - & 2 \\
\hline 
1351+640 & 0.088 & 2.5E+45 & 3.1E+45 & 5.2E+45 & 6 \\
\hline
\end{tabular}
\begin{minipage}{5.5truein}
1: Luminosity over indicated bandpass, assuming $H_o$=50 km s$^{-1}$ Mpc$^{-1}$
and $q_o$=0 (erg s$^{-1}$) \\
2: Probably a gravitationally lensed source (\cite{Omont96})
\end{minipage}
\end{table*}

The IR SED is smooth and indicates a wide and probably continuous
range of temperatures: 40--500K. In a purely thermal scenario this
smooth continuum can be explained by
a combination of cool host galaxy dust, warm ($\sim 100$K) starburst-related
dust (e.g.\ \cite{RR95}) and hot (T$_{eff} \sim 800-1300$K)
AGN-related dust (e.g.\ from a torus, \cite{PK93}). 
Alternatively the emission could originate in 
AGN-related dust which is sufficiently optically thick and/or geometrically
arranged (e.g.\ the warped disk model of \cite{Sand89} or a patchy torus) to
produce comparable emission over the wide range of observed
temperatures. This latter scenario is supported by the results of van
Bemmel \& Barthel (this volume) which demonstrate that RLQs have
significantly brighter far-IR emission than radio galaxies in samples
which were matched
so that the only difference should be one of orientation.

We also note that, with these data alone, we cannot rule out a significant
contribution from a non-thermal component in the mid- and near-IR.

\section{HIGH REDSHIFT QUASARS}

\begin{figure}[h]
    \leavevmode
\vskip -0.2in
  \centerline{\epsfig{file=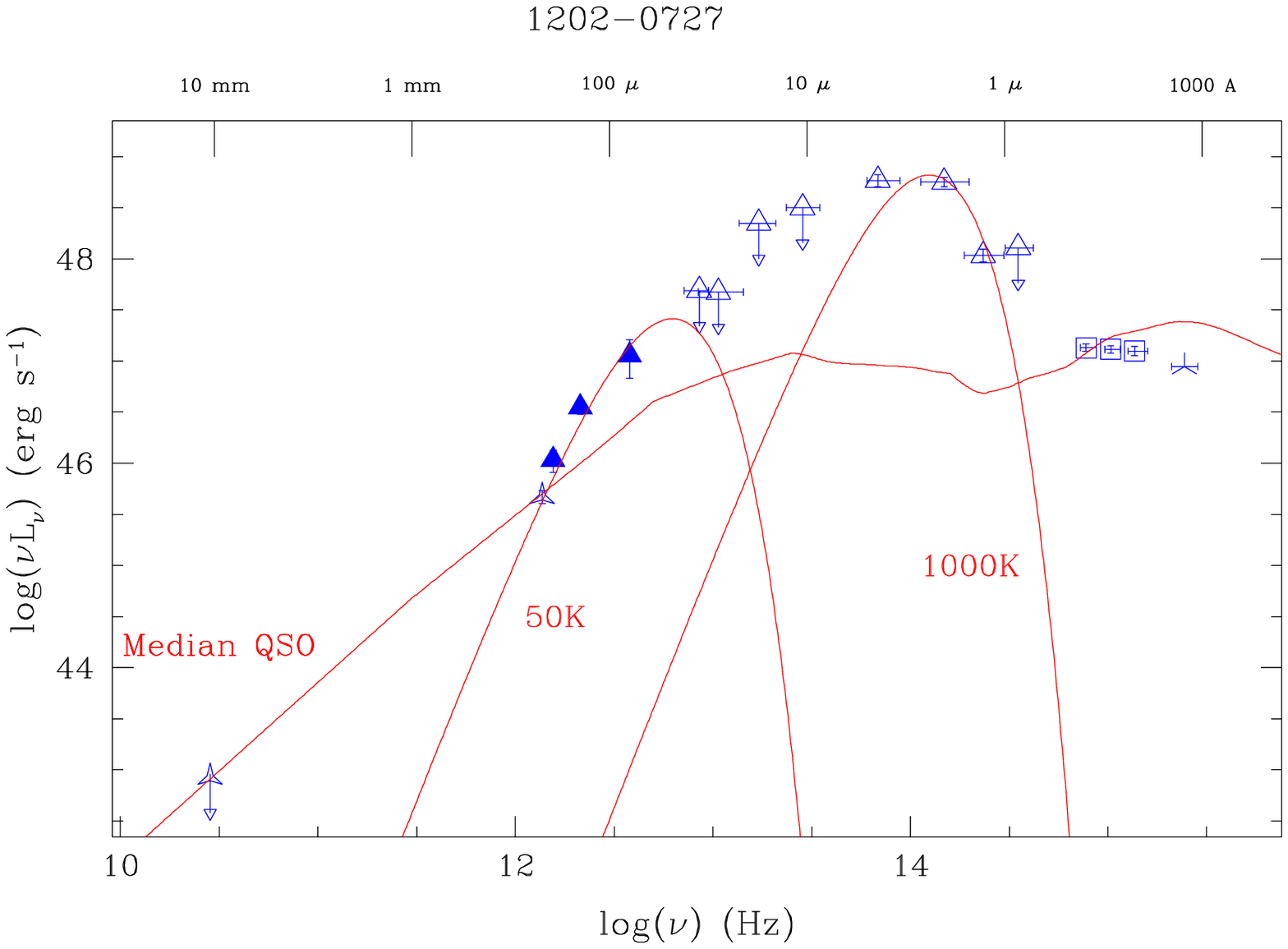,height=3.5truein}}
\vskip -1.2in
  \centerline{\epsfig{file=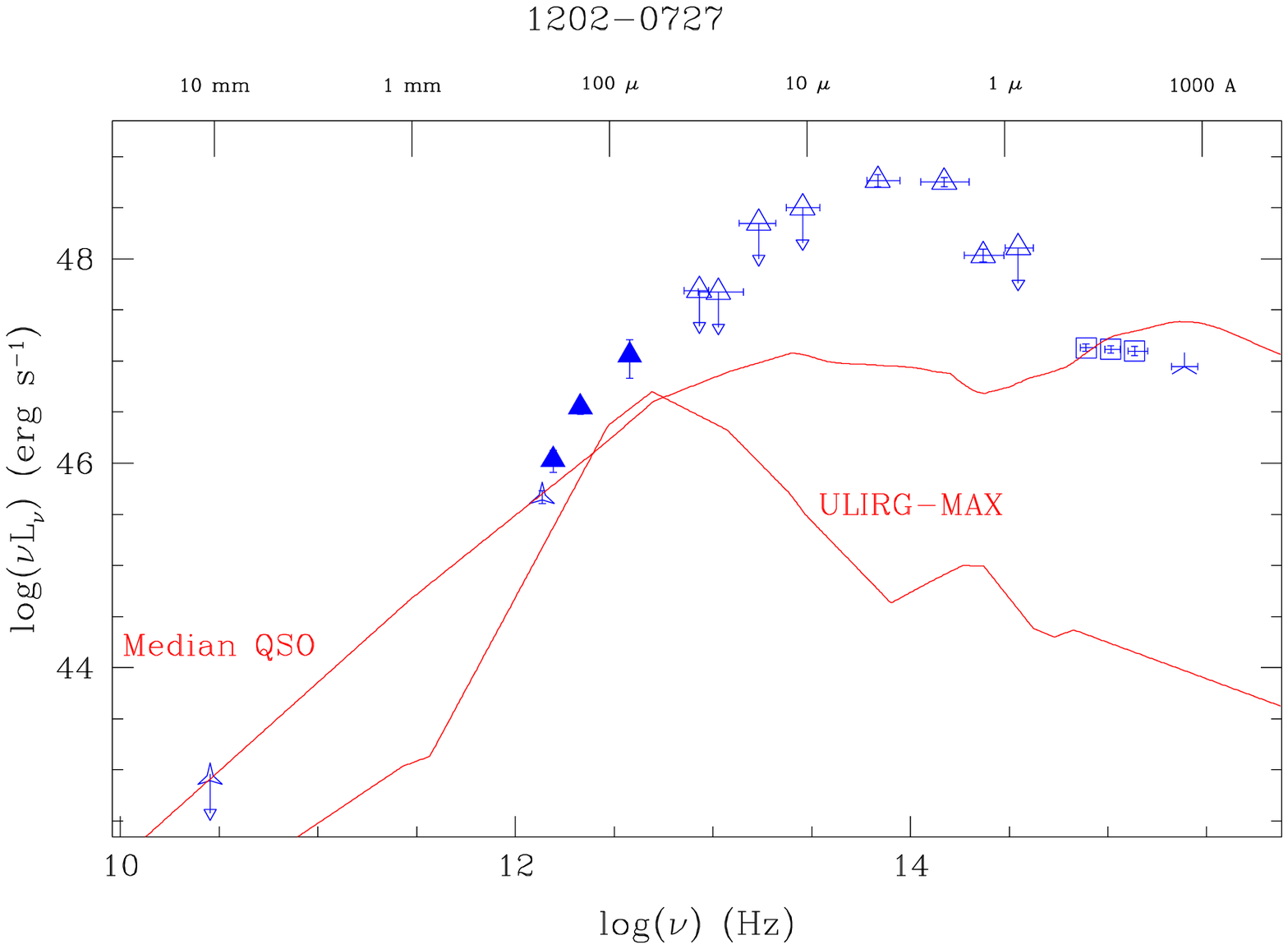,height=3.5truein}}
\vskip -1.1in
  \caption{\em The rest-frame, radio--UV SED of the high-redshift, RQQ
BR 1202$-$0727 including the ISO data points (open triangles). The
median SED for low redshift RQQs (\protect\cite{ewm94}), normalised to
the optical SED, is superposed
for comparison. {\bf Upper}: 
grey-body curves (assuming $\beta$=2)
at two extreme temperatures are superposed
{\bf Lower}: the SED of an ULIRG is superposed,
normalised to the highest H-band luminosity host galaxy observed in low redshift
sources (\protect\cite{McLeod97}).}
  \label{fig:1202}
\end{figure}

Our sample includes 9 quasars with redshift $> 0.9$. The 7 for which we have
reliable ISO detections are listed in Table~\ref{tab:lum} along with their
redshifts, preliminary estimates of several broad-band IR luminosities
and the number of ISO detections used in this determination.
The luminosities were determined via linear interpolation between the
data points and are not tabulated where there were insufficient
data to provide useful constraints. The IR-band luminosities,
determined assuming $H_o$=50 km s$^{-1}$ Mpc$^{-1}$
and $q_o$=0, are high,
ranging from 10$^{46-49}$ erg s$^{-1}$ ($\sim 10^{13-16} {\rm L_{\odot}}$).
For comparison, the low-redshift RQQ,
PG1351+640 (Table~\ref{tab:lum}) has luminosities $\sim 10^{45}$ erg s$^{-1}$
($\sim 10^{12} {\rm L_{\odot}}$).

The rest-frame, radio--UV SED of BR 1202$-$0727, the object with the
highest luminosity, L(2--200\,$\mu$m)
$\sim 10^{49}$ erg s$^{-1}$, is displayed in
Figure~\ref{fig:1202}.
Our ISO fluxes and upper limits are shown as open triangles
and our ground-based near-IR photometry,
obtained at the Steward Observatory 61$"$ telescope in March 1996, as open
squares. Multiwavelength data are from~\cite{Isaak94} and
\cite{McMahon94}. The median, low-redshift, RQQ SED (\cite{ewm94}), normalised
at $\sim 2$\,$\mu$m, is superposed for comparison.
Relative to the rest-frame optical emission, the IR emission in this
object is $\sim 2$ orders of magnitude stronger than is typical.
This is a complex source. It is a luminous (dust
$\sim 10^{11}{\rm M_{\odot}}$, \cite{Ohta96}), double (\cite{Omont96}),
CO source with a Ly$\alpha$ emission line companion (\cite{Hu96}).
Given its unprecendentedly strong mm and IR luminosities and double
nature it seems highly likely that the source is graviationally lensed
and so amplified by $\sim \times 10$. However lensing
cannot explain the unusual SED of BR 1202-0727, since partial lensing would
boost the smaller, optical/UV source rather than the
probably extended IR emitting region. 

Two extreme temperature grey-body curves (assuming an emissivity index,
$\beta =2$) are also superposed in Figure~\ref{fig:1202} (upper)
for illustration. Again a wide range of temperature is present, $50-1000$K
although the upper limits in the mid-IR prevent any constraint on the
emission at intermediate temperatures in this source. 
In Figure~\ref{fig:1202} (lower) an SED representing a ULIRG with the
maximum H-band host galaxy luminosity based on low redshift quasars (\cite{McLeod97})
is superposed. This comparison
illustrates that the far-IR emission is comparable or in excess of that
expected from a ULIRG and far exceeds that expected from the host galaxy 
(see also \cite{Wilkes97}).

\begin{figure}[h]
    \leavevmode
\vskip -0.2in
  \centerline{\epsfig{file=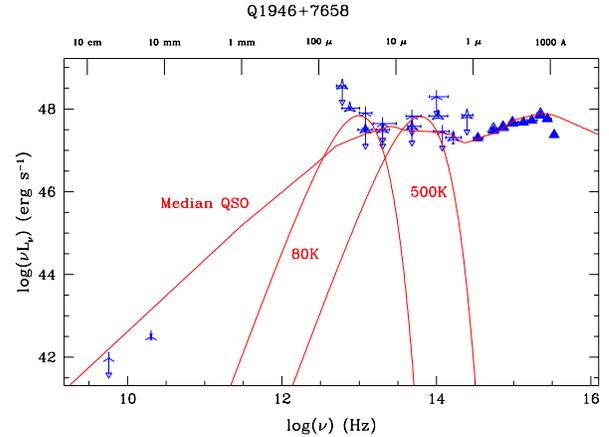,height=3.5truein}}
\vskip -1.1in
  \caption{\em The rest-frame, radio--UV SED of the high-redshift, RQQ
HS 1946+7658 including the ISO data points shown as open triangles.
The median SED for low redshift RQQs (\protect\cite{ewm94}) is superposed
for comparison. Grey-body curves (assuming $\beta$=2)
at two temperatures are superposed to illustrate the
range of temperature present in a pure thermal emission scenario.}
  \label{fig:1946}
\end{figure}

The rest-frame, radio--UV SED of HS 1946+7658 is displayed in
Figure~\ref{fig:1946}.
Our ISO fluxes and upper limits are shown as open triangles and the
remainder of the multiwavelength data were taken
from~\cite{Kuhn94} and~\cite{Kuhn96}. 
We also superpose grey-body curves (assuming dust with emissivity index,
$\beta =2$) at two extreme temperatures to illustrate the range of
temperature present in the data. The SED of HS 1946+7658
is remarkably similar to the low redshift, RQQ median, extending the
conclusion of~\cite{Kuhn94} based on the optical SED for this source.
As with the low redshift RQQ
PG1351+640, the IR continuum appears smooth. 
In a pure thermal scenario, this suggests the presence of dust covering a
continuous range from $\sim <80-500$K rather than
two distinct low and high temperature components.
The low temperature limit is determined by the longest
wavelength at which we have data rather than a real cut-off in the
source. 

In summary, the broad-band IR luminosities, ranging
from 10$^{46-48}$ erg s$^{-1}$ ($\sim 10^{13-15} {\rm L_{\odot}}$, assuming
BR 1202-0727 is lensed), 
are $\sim 2$ orders of magnitude
higher than those of Ultraluminous IR galaxies (ULIRGs) whose SEDs peak
in the IR with luminosities $\sim 10^{11-13} {\rm L_{\odot}}$ (e.g.\
\cite{Sand96}), even allowing for the different assumed Hubble
constant (\cite{Sand96} use $H_o = 75$ km s$^{-1}$ Mpc$^{-1}$).
In the far-IR these luminosities are too high to be due to the
host galaxy. 
As for the low redshift sources (\ref{sec:1351}), the inferred temperature
ranges are generally too broad to be fit by standard, AGN 
molecular torus models (\cite{PK93}), which can reproduce the
higher temperature emission. Thus more complex models or a strong starburst
component are required to provide the lower temperature emission.

\section*{ACKNOWLEDGMENTS}
We wish to thank the staff at the ISOPHOT data center in Heidelberg, in particular
Martin Haas, for their help and hospitality while working on these data.
We also thank the ISO staff at IPAC for their continual support and help
both during visits and remotely. This work is supported by NASA grant
NAGW-3134.


\begin{thebibliography}{}

\bibitem[\protect\astroncite{Barthel}{(1999)}]{Barthel99}
Barthel, P. 1999 in ``Quasars and Cosmology", Gary Ferland and Jack
Baldwin, Eds, ASP Conf. Proc. {\it in press}
\bibitem[\protect\astroncite{Elvis et al.}{(1994)}]{ewm94}
Elvis,~M.~S., Wilkes,~B.~J., McDowell,~J.~C., Green,~R.~F.,
Bechtold,~J., Willner,~S.~P., Cutri,~R., Oey,~M,~S., \& Polomski,~E.
1994, ApJS, 95, 1 
\bibitem[\protect\astroncite{Gabriel et al.}{(1997)}]{Gabriel97}
Gabriel, C. et al. 1997 in Proceedings of ADASS VI, ASP Conf. Ser. 125,
G. Hunt \& H.E. Payne, eds., [ASP], p. 108
\bibitem[\protect\astroncite{Gear et al.}{(1994)}]{Gear94}
Gear, W., Stevens, J.A., Hughes, D.H., Litchfiueld, S.J.,
Robson, E.I.,
Terasanta, H., Valtaoja, E., Steppe, H., Aller, M.F., \& Aller, H.D. 1994,
MNRAS, 267, 167
\bibitem[\protect\astroncite{Herbstmeier et al.}{(1998)}]{herb98}
Herbstmeier, U., Abraham, P., Lemke, D., Laureijs, R. J.,
Klaas, U., Mattila, K., Leinert, C., Surace, C. \& Kunkel, M.
1998, AA, 332, 739
\bibitem[\protect\astroncite{Haas et al.}{(1998)}]{Haas98}
Haas,~M., Chini,~R., Meisenheimer,~K., Stickel,~M.,
Lemke,~D., Klaas,~U. \& Kreysa,~E. 1998 ApJL, 503, 109
\bibitem[\protect\astroncite{Hu et al.}{(1996)}]{Hu96}
Hu, E.M., McMahon, R.G. \& Egami, E. 1996, ApJ, 459, L53
\bibitem[\protect\astroncite{Isaak et al.}{(1994)}]{Isaak94}
Isaak, K.G., McMahon, R.G., Hills, R.E. \& Withington, S. 1994, MNRAS,
269, L28
\bibitem[\protect\astroncite{Kuhn}{(1996)}]{Kuhn96}
Kuhn, O. 1996, PhD Thesis, Harvard University
\bibitem[\protect\astroncite{Kuhn et al.}{(1994)}]{Kuhn94}
Kuhn, O., Bechtold, J., Cutri, R., Elvis, M. \& Reike, M. 1994,
ApJ, 438, 643
\bibitem[\protect\astroncite{Lemke et al.}{(1996)}]{phot96}
Lemke, D. \etal\ 1996, AA 315, L64
\bibitem[\protect\astroncite{McLeod}{(1997)}]{McLeod97}
Mcleod, K.K. 1997 in ``Quasar Hosts", D.L. Clements \& I.
Perez-Fournon eds., [Springer], p. 45
\bibitem[\protect\astroncite{McMahon et al.}{(1994)}]{McMahon94}
McMahon, R.G., Omont, A., Bergeron, J., Kreysa, E. \& Haslam, C.G.T.
1994, MNRAS, 267, L9
\bibitem[\protect\astroncite{Ohta et al.}{(1996)}]{Ohta96}
Ohta, K., Yamada, T., Nakanishi, K., Kohno, K., Akiyama, M. \& Kawabe,
R. 1996, Nature, 382, 426
\bibitem[\protect\astroncite{Omont et al.}{(1996)}]{Omont96}
Omont, A., Petitjean, P., Guilloteau, S., McMahon, R.G., Solomon, P.M.
\& Pecontal, E. 1996, Nature, 382, 428
\bibitem[\protect\astroncite{Pier \& Krolik}{(1993)}]{PK93}
Pier, E.A. \& Krolik, J.H. 1993, ApJ, 418, 673
\bibitem[\protect\astroncite{Rowan-Robinson}{(1995)}]{RR95}
Rowan-Robinson, M. 1995, MNRAS, 272, 737
\bibitem[\protect\astroncite{Sanders \& Mirabel}{(1996)}]{Sand96}
Sanders, D.B. \& Mirabel, I.F. 1996, ANRAA, 34, 749
\bibitem[\protect\astroncite{Sanders et al.}{(1989)}]{Sand89}
Sanders,~D., Phinney,~E.S., Neugebauer,~G.,
Soifer,~B.T., \& Matthews,~K. 1989, ApJ, 347, 29
\bibitem[\protect\astroncite{Wilkes}{(1999)}]{Wilkes99}
Wilkes, B.J. 1999 in ``Quasars and Cosmology", Gary Ferland and Jack
Baldwin, Eds, ASP Conf. Proc. {\it in press}
\bibitem[\protect\astroncite{Wilkes}{(1997)}]{Wilkes97}
Wilkes, B.J. 1997 in ``Quasar Hosts", D.L. Clements \& I.
Perez-Fournon eds., [Springer], p. 136
\end{thebibliography}
\end{document}